\documentclass[conference]{IEEEtran}
\IEEEoverridecommandlockouts
\usepackage{cite}
\usepackage{amsmath,amssymb,amsfonts}
\usepackage{algorithmic}
\usepackage{graphicx}
\usepackage{textcomp}
\usepackage{xcolor}
\usepackage[switch,mathlines]{lineno}
\def\BibTeX{{\rm B\kern-.05em{\sc i\kern-.025em b}\kern-.08em
    T\kern-.1667em\lower.7ex\hbox{E}\kern-.125emX}}

\usepackage{booktabs,array,colortbl,tabularx}
\usepackage{xcolor}
\usepackage{pifont}
\usepackage[bookmarks=false]{hyperref}
\usepackage{booktabs}
\usepackage{microtype}
\usepackage{xspace}
\usepackage{enumitem}
\usepackage{balance}
\usepackage{stfloats}
\usepackage{colortbl}
\usepackage{multicol}
\usepackage{multirow}
\usepackage{pifont}
\usepackage{algorithm}
\usepackage{booktabs}

\begin{document}

\title{\textit{AceVLN}: Asynchronous Cloud–Edge Deployment of VLNs via Latent Scheduling and Compensation}
\author{
    \IEEEauthorblockN{
        Zihao Zheng$^{1,*}$,
        Hangyu Cao$^{1,*}$,
        Yulong Han$^{1}$,
        Xichang Wu$^{2}$,
        Zhihao Mao$^{1}$,
        Maoliang Li$^{1}$,\\
        Jiayu Chen$^{1}$,
        Hailong Zou$^{1}$,
        Donggang Cao$^{1}$,
        Xuanzhe Liu$^{1}$,
        Hong Mei$^{1}$,
        Xiang Chen$^{1,\dagger}$
    }
    \IEEEauthorblockA{
        $^{1}$ Peking University
        \quad
        $^{2}$ Beijing Jiaotong University\\
        $^{*}$ Equal Contribution
        \quad
        $^{\dagger}$ Corresponding author: xiang.chen@pku.edu.cn
    }
}
\maketitle

% ==================================================================
% tex
% ==================================================================
\begin{abstract}
Cloud--edge deployment enables asynchronous dual-system VLN on resource-constrained robots, but inference and communication delays cause latent mismatch and navigation errors. 
Existing adaptation methods reuse historical guidance but cannot provide updated navigation information when it becomes inadequate. 
We propose \textit{AceVLN}, a training-free framework for active latent refresh and edge-side compensation. 
Our analysis reveals that action-relevant mismatch is concentrated in a compact Critical Subspace and can be tracked through fast-system feature drift. 
Critical Feature Drift Entropy reflects navigation sensitivity and guides adaptive refresh, while Critical Drift Continuity indicates when recent latent updates can support compensation during latency spikes. 
\textit{AceVLN} integrates subspace-decoupled mismatch modeling, entropy-guided refresh scheduling, and continuity-guided compensation. 
Experiments show that \textit{AceVLN} improves success rates by 1.7--2.4 percentage points over the strongest baselines, while reducing amortized computation time by 2.7--10.0\% and FLOPs by 18.4--22.3\%.
\end{abstract}
\section{\textbf{Introduction}}
\label{sec:introduction}

%%%%%%%%%%%%%%%%%%%%%%%%%
% 1. VLN 双系统 + 端云
%%%%%%%%%%%%%%%%%%%%%%%%%
Streaming Vision-Language Navigation (VLN) enables robots to generate real-time navigation actions from continuous visual observations and language instructions~\cite{navila,streamvln,zhang2025uninavid}. 
To balance high-level reasoning and real-time control, recent models adopt an asynchronous dual-system architecture, where a slow system performs low-frequency reasoning and a lightweight fast system generates high-frequency actions~\cite{dualvln,tic}. 
However, the substantial computational and memory demands of the slow system hinder its deployment on resource-constrained robots~\cite{survey}. Consequently, practical systems commonly adopt edge-cloud collaboration (ECC), deploying the slow system in the cloud and the fast system onboard to leverage cloud computing resources while preserving real-time local control~\cite{cloudedge, asyncvla, roboecc}.

%%%%%%%%%%%%%%%%%%%%%%%%%
% 2. Latent 背景
%%%%%%%%%%%%%%%%%%%%%%%%%
In dual-system VLN, the latent serves as an intermediate representation that transfers multi-step navigation guidance from cloud-side reasoning to edge-side control. 
Specifically, the slow system encodes uploaded visual observations and navigation instructions into a latent containing guidance for subsequent navigation steps, which the fast system combines with real-time observations to generate actions~\cite{havana,cloudedge}. 
However, cloud inference and communication delays cause the robot to continuously move while the fast system reuses a latent generated from historical observations~\cite{sparkvln}. 
As the robot's position and navigation requirements change, the historical latent becomes increasingly inconsistent with the ideal latent that would be generated from current observations, resulting in \textit{\underline{latent mismatch}}. 
Its severity depends on both edge-cloud latency and the rate of navigation-state changes, causing outdated guidance to induce action deviations and accumulated trajectory errors, ultimately degrading navigation accuracy.

%%%%%%%%%%%%%%%%%%%%%%%%%
% 3. SOTA + Challenge
%%%%%%%%%%%%%%%%%%%%%%%%%
Existing methods mitigate latent mismatch through adaptation. AsyncVLA~\cite{asyncvla} employs an onboard Edge Adapter to refine actions using real-time observations, while TIC-VLA~\cite{tic} incorporates delayed semantic features and latency metadata into action generation. CloudEdgeVLA~\cite{cloudedge} further improves tolerance to stale guidance through training with paired fresh and delayed observations. However, these methods primarily improve action or representation-level adaptation to historical guidance, rather than explicitly assessing action-relevant mismatch to schedule new latent generation or compensate during delayed updates. When historical guidance no longer reflects current navigation requirements, adaptation alone cannot provide updated guidance. 
\textbf{\textit{These limitations motivate Active Latent Refresh, which proactively requests updated navigation guidance based on runtime mismatch.}}

\begin{figure}[!t]
    \centering
    \includegraphics[width=3.3in]{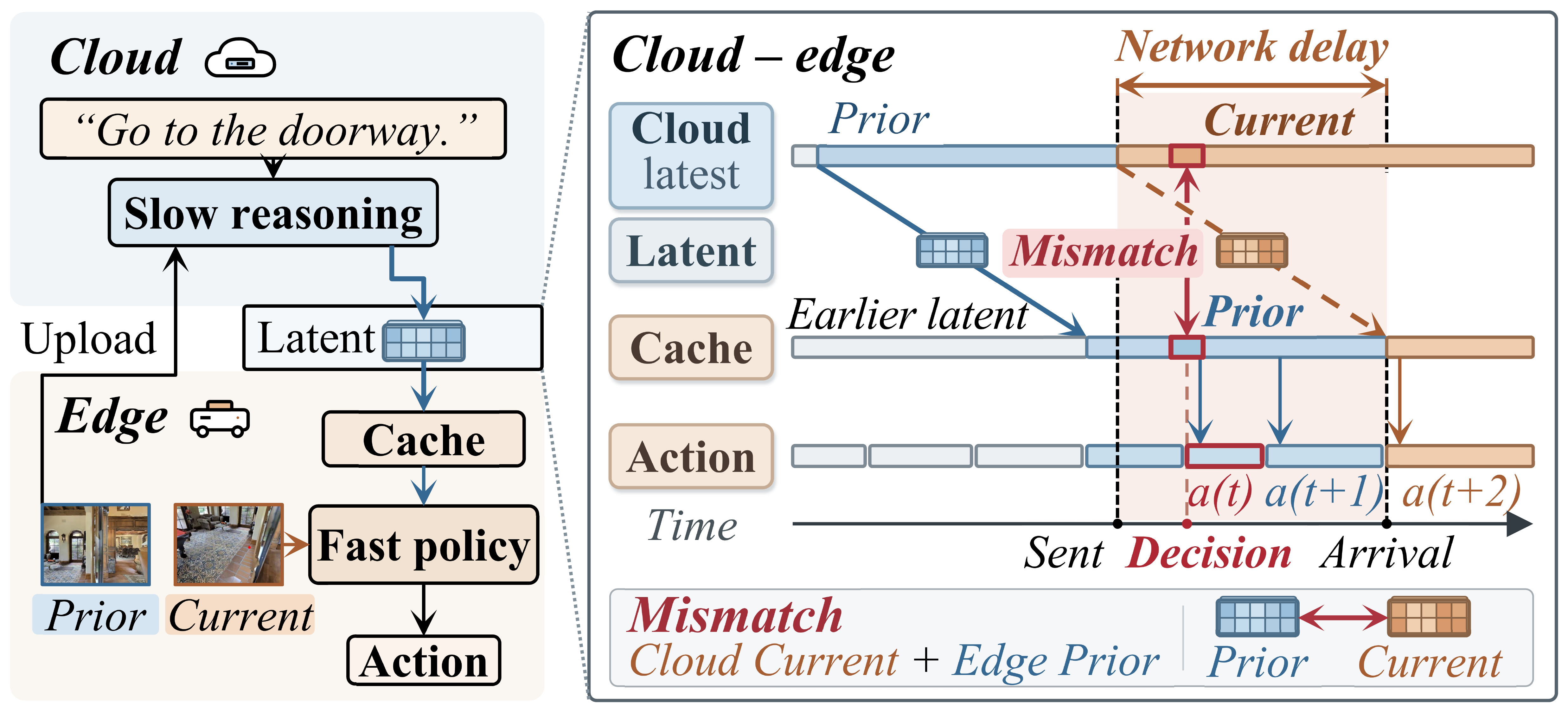}
    \vspace{-2mm}
    \caption{Latent Mismatch in Edge-Cloud Collaborative VLN.}
    \label{fig:latent}
    \vspace{-8mm}
\end{figure}

%%%%%%%%%%%%%%%%%%%%%%%%%
% 4. Challenge
%%%%%%%%%%%%%%%%%%%%%%%%%
However, achieving efficient and reliable refresh requires addressing three key challenges:
\textit{Challenge \ding{172}: Latent Mismatch Is Hard to Model.}
A single latent entangles guidance for multiple navigation steps, whose mismatches have different impacts on current actions. Consequently, overall latent differences conflate action-critical deviations with less relevant future-step changes, obscuring their actual impact on current actions.
\textit{Challenge \ding{173}: Latent Refresh Timing Is Hard to Determine.}
Different navigation states have varying tolerances to action deviations. Similar deviations can lead to substantially different navigation errors, making fixed refresh intervals prone to unnecessary cloud updates or delayed refreshes that degrade navigation accuracy.
\textit{Challenge \ding{174}: Latent Mismatch Is Hard to Handle under Latency Spikes.}
Even when refresh timing is accurately determined, unexpected latency spikes can delay updated latents~\cite{actecc,yang2026asyncshield}, leaving the edge system without fresh navigation guidance as a reference for tracking and correcting mismatch during the refresh wait.

%%%%%%%%%%%%%%%%%%%%%%%%%
% 5. Analysis
%%%%%%%%%%%%%%%%%%%%%%%%%
To address these challenges, we analyze latent mismatch and its impact on actions, deriving three key insights:
\textit{Analysis \ding{172}:} 
We examine latent variations across subspaces, action deviations, and feature responses in the fast system, revealing that changes in a compact Critical Subspace dominate action deviations and can be tracked through fast-system features for lightweight runtime mismatch assessment.
\textit{Analysis \ding{173}:} 
We investigate error tolerance and directional drift distributions across navigation steps, revealing that higher Critical Feature Drift Entropy indicates greater navigation sensitivity, guiding dynamic latent refresh scheduling.
\textit{Analysis \ding{174}:}
We analyze critical drift evolution and latent update dynamics under delayed refreshes, revealing short-term continuity in latent updates under consistent drift directions, motivating adaptive compensation on the edge through extrapolation of consistent historical updates.

%%%%%%%%%%%%%%%%%%%%%%%%%
% 6. Method
%%%%%%%%%%%%%%%%%%%%%%%%%
Guided by these insights, we propose \textit{AceVLN}, a training-free framework for active latent refresh and compensation in cloud--edge streaming VLN.
\textit{Method \ding{172}:} Subspace-Decoupled Latent Mismatch Modeling isolates the Critical Subspace and tracks its mismatch through fused features of the fast system for lightweight runtime assessment.
\textit{Method \ding{173}:} Drift-Distribution-Guided Latent Refresh Scheduling combines critical feature drift with its directional entropy to adapt latent refresh timing to navigation sensitivity.
\textit{Method \ding{174}:} Continuity-Guided Latent Compensation extrapolates recent critical latent updates, using drift continuity and magnitude to control compensation scope and strength during pending refreshes and latency spikes.
Together, these mechanisms reduce cloud updates while preserving navigation accuracy.

%%%%%%%%%%%%%%%%%%%%%%%%%
% 7. Contribution
%%%%%%%%%%%%%%%%%%%%%%%%%
In summary, our contributions are threefold:
\begin{itemize}
    \item We reveal that latent mismatch affects actions primarily through a compact critical subspace, uncovering new insights into navigation error tolerance and temporal mismatch evolution.
    \item Based on these insights, we propose \textit{AceVLN}, a training-free framework that combines adaptive latent scheduling and complementary edge-side compensation through unified mismatch modeling, enabling efficient and reliable VLN deployment.
     \item On A100 with Jetson Orin/Thor, we quantify the trade-off between cloud calls, total compute, and edge response time. Relative to \emph{Bridge}, \textit{AceVLN} cuts FLOPs by 18.4\%--22.3\% and amortized compute by 2.7\%--10.0\%, with 5--13\,ms higher mean response time. Success rate improves by 1.7--2.4 percentage points over the best baselines, and compensation adds 2.7 points under latency spikes.
\end{itemize}

\section{\textbf{Background and Motivation}}
\label{sec:background}

% =======================================================
% subsection 1: Streaming Vision-Language Navigation Models
% =======================================================
\subsection{\textbf{Streaming Vision-Language Navigation Models}}
\label{sec:background_1}

\subsubsection{\textbf{Asynchronous Dual-System Architecture}}
Streaming VLN typically adopts an asynchronous dual-system architecture~\cite{dualvln,tic,tan2026vbotnav,chen2026latranav}.
A large VLM-based slow system periodically generates multi-step navigation guidance from instructions and visual observations, while a lightweight fast system combines this guidance with real-time observations to generate high-frequency actions.

\subsubsection{\textbf{Latent Transfer in Dual-System VLN}}
\begin{figure}[!t]
    \centering
    \includegraphics[width=3.3in]{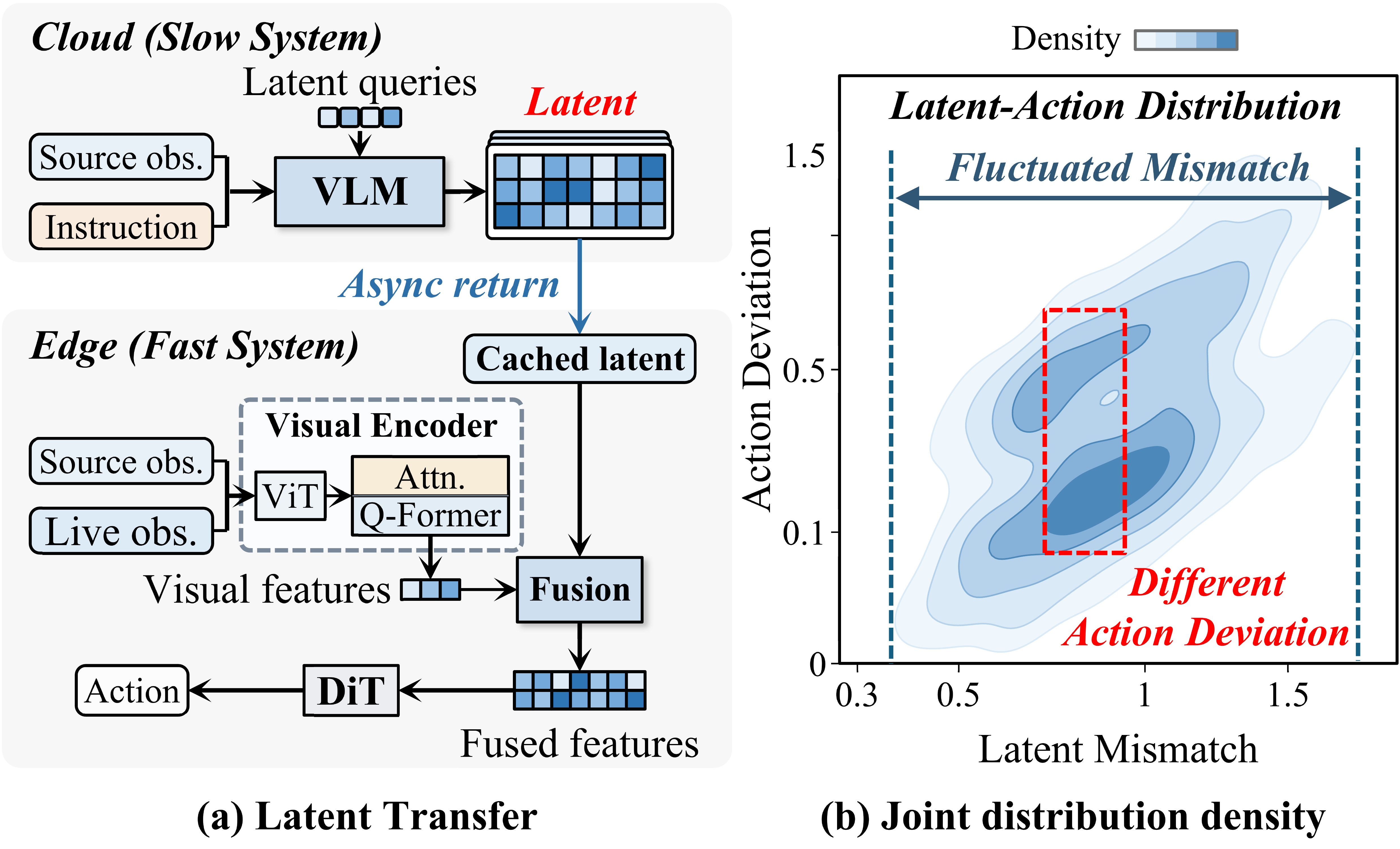}
    \vspace{-2mm}
    \caption{Latent Transfer and Mismatch-induced Action Deviation.}
    \label{fig:mismatch}
    \vspace{-6mm}
\end{figure}

As shown in Fig.~\ref{fig:mismatch}(a), the slow system encodes the instruction and source observation into a shared high-dimensional latent encapsulating multi-step guidance~\cite{zheng2026flight,zhu2026fsdvln}.
Across consecutive control steps, the fast system fuses encoded source and live observations with the cached latent into fused features to decode real-time actions.

% =======================================================
% subsection 2: ECC Deployment for Streaming VLN Models
% =======================================================
\subsection{\textbf{ECC Deployment for Streaming VLN Models}}
\label{sec:background_2}

\subsubsection{\textbf{ECC Deployment of Dual-System VLN}}
Given the computational and memory demands of the slow system~\cite{survey}, edge-cloud collaboration (ECC) can place high-level planning in the cloud and keep local control onboard~\cite{asyncvla,tan2026vbotnav,yang2026entitynav}.
The edge uploads observations and asynchronously receives updated latents, continuously generating actions from the latest cached latent~\cite{cloudedge}.

\subsubsection{\textbf{Latent Mismatch and Requirements of Active Refresh}}
Cloud inference and communication delays force the moving robot to reuse historical latents~\cite{li2026fastslowdriving,tic}, creating \textit{latent mismatch} with ideal latents for live observations and degrading navigation accuracy.
Adaptation methods~\cite{asyncvla,tic,cloudedge} adjust actions to tolerate stale guidance rather than updating it, motivating \textit{Active Latent Refresh} at extra cloud computation and communication costs~\cite{pohland2026offload,chinchali2019offloading}.
However, existing uncertainty triggers~\cite{yang2026tud} or latent predictors~\cite{liu2026latentbridge} treat the latent monolithically, without modeling how mismatch actually impacts ongoing navigation.

\subsubsection{\textbf{Motivation for Latent Mismatch Analysis}}
AdaNav uses action uncertainty to decide when to reason~\cite{adanav}, while ASPEN predicts whether the previous navigation command can be reused~\cite{akustia2026aspen}. For latent refresh, the question is how much reusing historical guidance changes the current action.
Comparing historical and ideal latents under identical live observations (Fig.~\ref{fig:mismatch}(b)) reveals that mismatch fluctuates widely (\textit{Fluctuated Mismatch}), yet similar latent deviations yield markedly \textit{Different Action Deviations}—with large shifts often causing minimal action error.
Because one latent entangles multi-step guidance, overall variations obscure current-step control errors; meanwhile, ideal latents are unavailable on the edge at runtime, state error tolerances vary, and latency spikes unpredictably delay updates.
While rendering coarse mismatch metrics ineffective, this decoupling reveals substantial headroom to avoid costly updates—motivating us to dissect how mismatch propagates through edge features, state tolerance, and delayed updates.
\section{\textbf{Latent Mismatch Analysis}}
\label{sec:analysis}

In asynchronous edge--cloud VLN, the fast system acts on a cached latent between cloud updates. We examine which components of latent mismatch affect the action and can be tracked on the edge, when to request a fresh latent, and when extending the last update reduces mismatch during the wait.

\begin{figure}[!t]
    \centering
    \includegraphics[width=3.3in]{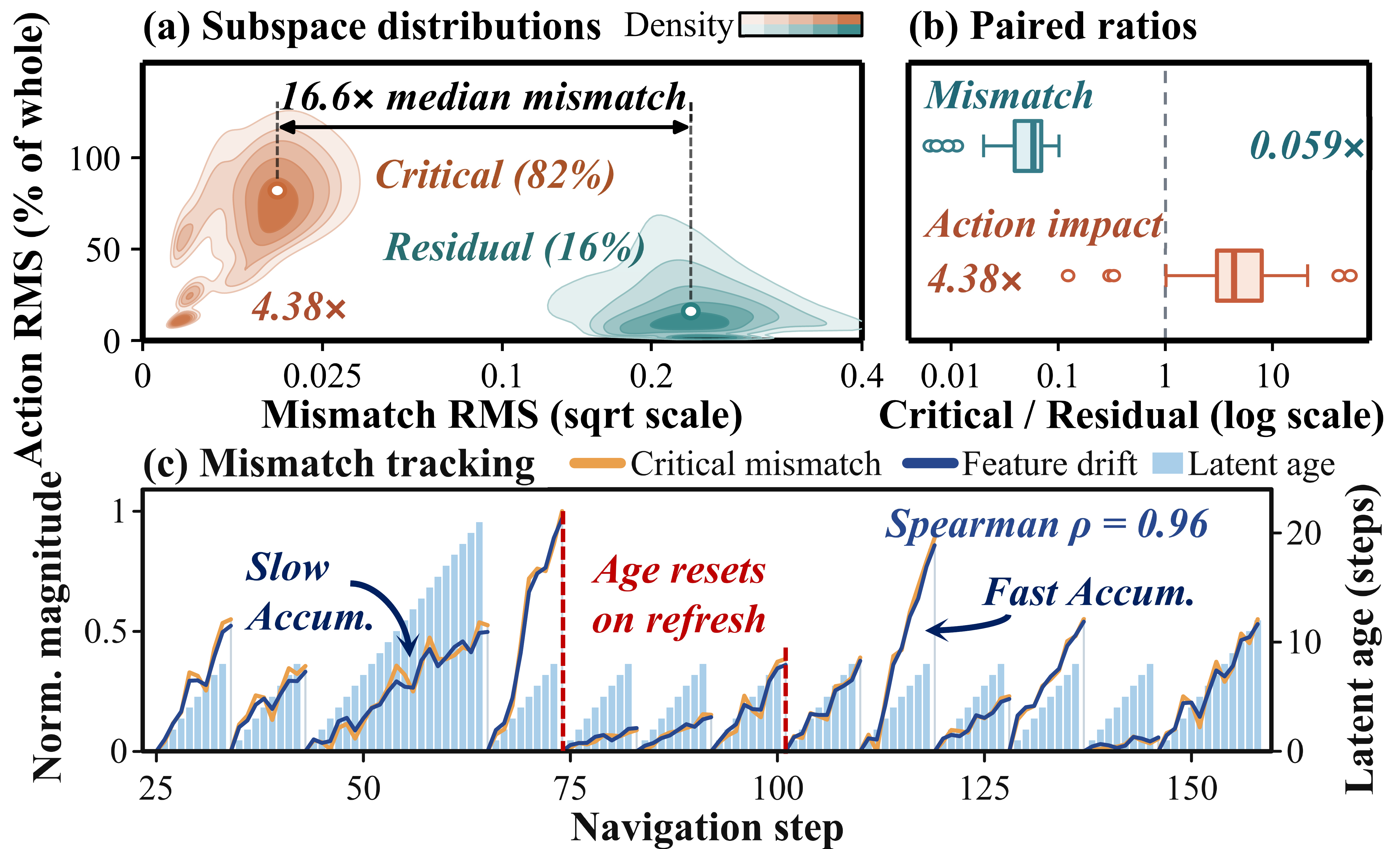}
    \vspace{-2mm}
    \caption{Subspace Decomposition and Critical Mismatch Tracking.}
    \label{fig:analysis1}
    \vspace{-6mm}
\end{figure}

% ========================================================================
% subsection 1: Subspace Decomposition for Latent Mismatch Modeling
% ========================================================================
\subsection{\textbf{Subspace Decomposition for Latent Mismatch Modeling}}
\label{sec:analysis_1}

To understand how latent mismatch affects the current navigation action, we compare the trajectories predicted by the fast system under the same observation using either the cached latent or a fresh latent. Fig.~\ref{fig:mismatch}(b) shows that even a large latent difference can induce a small action deviation. Since a latent encodes multi-step guidance, we ask which directions of mismatch affect the current prediction.

At each refresh boundary, we fix the observation $o$ and evaluate the trajectory Jacobian $J_a=\partial\hat{\mathbf a}(o,z)/\partial z$ at the cached latent $z_k$. In its singular value decomposition $J_a=U\Sigma V^\top$, the two leading right singular vectors form the columns of the local basis $V_c^{\rm loc}$ and span the \emph{Critical Subspace}. Their orthogonal complement forms the \emph{Residual Subspace}. For the fresh latent $z^\star$ at that same observation, we decompose the difference $\Delta z=z^\star-z_k$ as
\begin{equation}
    \Delta z_c = V_c^{\rm loc}(V_c^{\rm loc})^\top \Delta z, \qquad
    \Delta z_r = \Delta z - \Delta z_c .
    \label{eq:analysis_latent_components}
\end{equation}
We add each component separately to $z_k$ at 25\% of its measured magnitude and compare the resulting trajectory with the prediction under $z_k$, holding the observation and sampling seed fixed. Across 38 refresh boundaries from five trajectories, the median critical-to-residual ratio is 0.059 for mismatch RMS and 4.38 for induced action RMS (Fig.~\ref{fig:analysis1}(a,b)). Thus, the residual component can dominate the latent distance even when the smaller critical component has a greater effect on the current action.

Measuring critical mismatch directly requires the fresh latent, which is unavailable on the edge during the refresh wait. We use the fast system's fused features as a runtime proxy and identify feature directions that respond to critical latent perturbations. The map $P_c$ expresses feature changes in these $r$ response coordinates (Sec.~\ref{tex:method_1}). Using the same cached latent $z_k$, we obtain fused features $h_s=h(o_{s_k},z_k)$ at its source observation and $h_t=h(o_t,z_k)$ at the current observation. We define \emph{Critical Subspace Feature Drift} as
\begin{equation}
    \mathbf{d}_t = P_c(h_t - h_s), \qquad D_t = \|\mathbf{d}_t\|_2,
    \label{eq:analysis_drift}
\end{equation}
where $d_{t,i}$ is the signed change along feature response direction $i$. This drift captures observation-induced feature changes under a fixed latent. In Fig.~\ref{fig:analysis1}(c), latent age increases uniformly between refreshes, while critical mismatch accumulates at different rates. $D_t$ tracks slow and rapid changes, with Spearman $\rho=0.96$ against offline critical mismatch over the plotted steps.

\noindent\textit{\textbf{Insight \ding{172}:} Small perturbations along a few critical directions can have a large effect on the current action. Their fused-feature responses provide coordinates for tracking action-relevant mismatch on the edge.}

\begin{figure}[!t]
    \centering
    \includegraphics[width=3.3in]{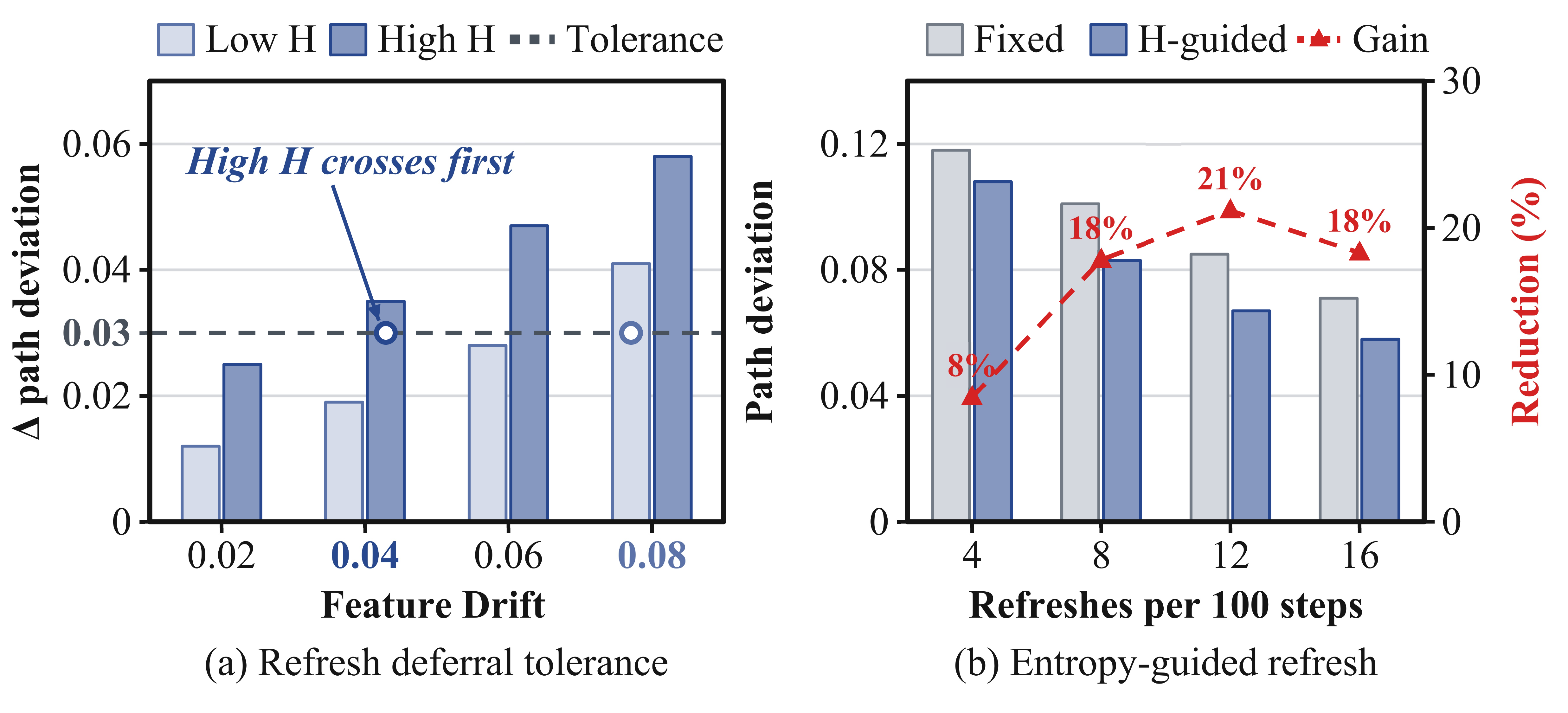}
    \vspace{-2mm}
    \caption{Refresh deferral tolerance and entropy-guided scheduling gains.}
    \label{fig:analysis2}
    \vspace{-6mm}
\end{figure}

% ========================================================================
% subsection 2: Tolerance Margin Analysis for Refresh Timing
% ========================================================================
\subsection{\textbf{Tolerance Margin Analysis for Refresh Timing}}
\label{sec:analysis_2}

Critical feature drift gives an estimate of mismatch, and similar drift magnitudes can have different consequences for the subsequent path. To measure a state's tolerance for delaying refresh, we start from the same navigation state and either request a fresh latent immediately or defer the request by a fixed number of steps. With edge--cloud latency held constant, the cost of deferral is the additional path deviation over the same horizon, $\Delta e_{\mathrm{path}}=e_{\mathrm{defer}}-e_{\mathrm{now}}$.

We examine whether the distribution of critical feature drift at the initial decision step distinguishes states with different tolerances. For $D_t>0$, we define \emph{Critical Feature Drift Entropy} over the $r$ feature response directions as
\begin{equation}
    H_t = -\sum_{i=1}^{r} p_{t,i}\log p_{t,i}, \quad
    p_{t,i} = \frac{d_{t,i}^2}{\sum_{j=1}^{r} d_{t,j}^2},
    \label{eq:critical_entropy}
\end{equation}
where $p_{t,i}$ is the share of drift energy in direction $i$ and $0\log 0=0$. Higher entropy indicates that the drift is spread more evenly across directions.

Fig.~\ref{fig:analysis2}(a) compares states with high and low entropy at the same initial drift magnitude $D_t$. At $D_t=0.04$, deferring the request raises $\Delta e_{\mathrm{path}}$ above the path-deviation tolerance of 0.03 in the high-entropy group. The low-entropy group stays within tolerance at that same drift magnitude and first exceeds it at the sampled value of $D_t=0.08$. This difference motivates lowering the refresh threshold on $D_t$ when $H_t$ is high. At a matched budget of 12 requests per 100 steps, the entropy-guided threshold reduces path deviation by 21\% relative to the fixed threshold (Fig.~\ref{fig:analysis2}(b)).

\begin{figure}[!t]
    \centering
    \includegraphics[width=3.3in]{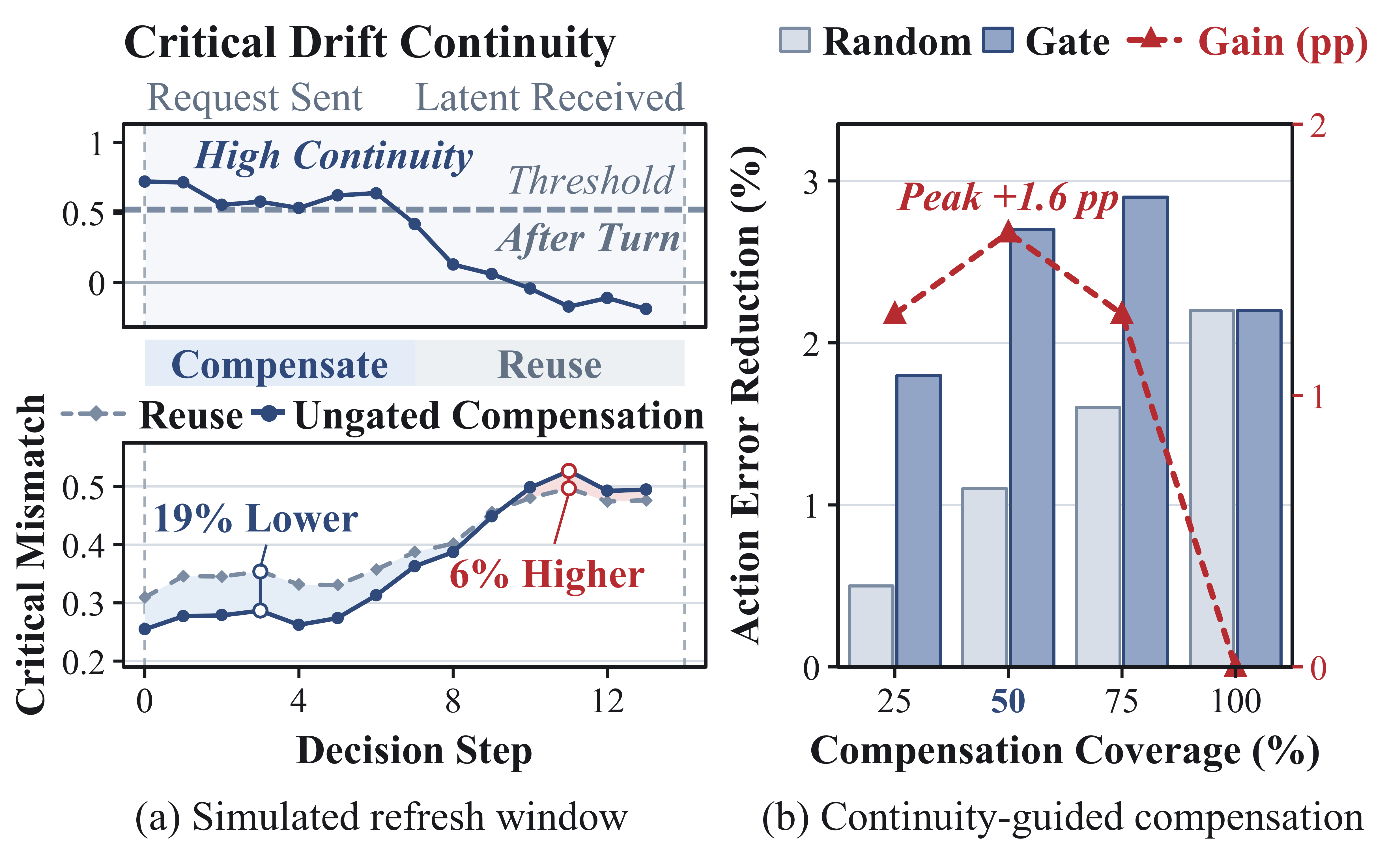}
    \vspace{-2mm}
    \caption{Critical drift continuity and latent compensation effectiveness.}
    \label{fig:analysis3}
    \vspace{-4mm}
\end{figure}

\begin{figure*}[!t]
    \centering
    \includegraphics[width=7in]{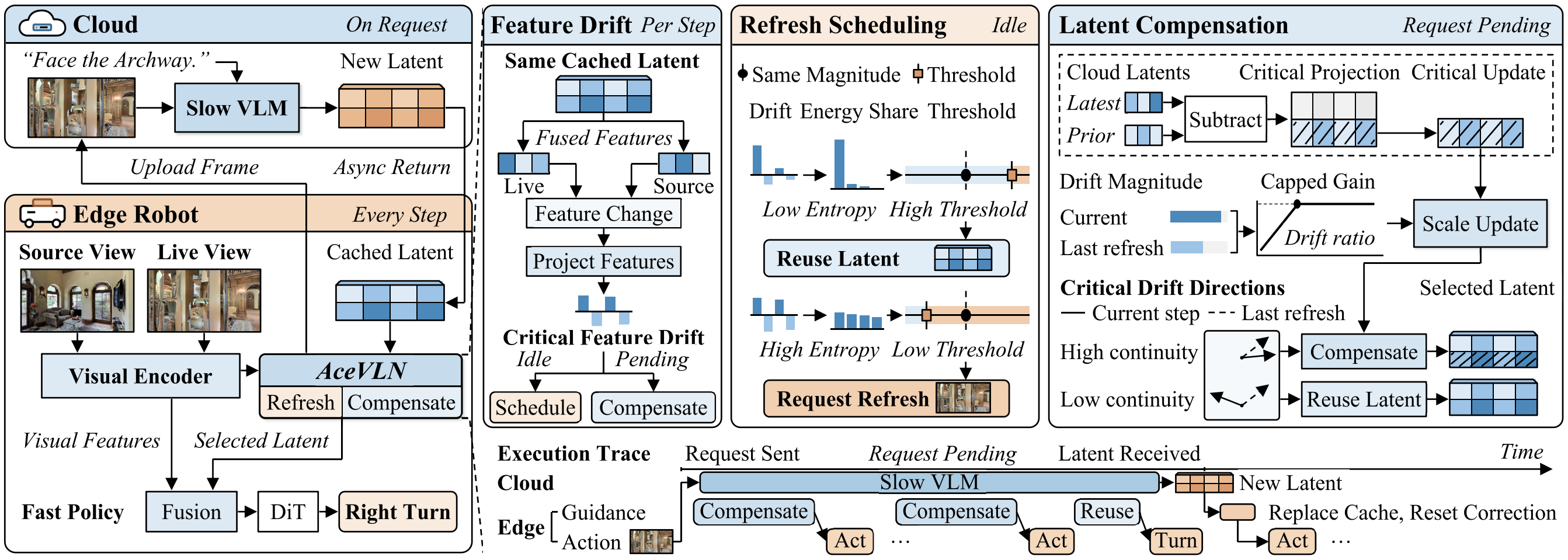}
    \vspace{-2mm}
    \caption{Overview of the \textit{AceVLN} Framework.}
    \vspace{-4mm}
    \label{fig:framework}
    \label{fig:placeholder-double}
\end{figure*}

\noindent\textit{\textbf{Insight \ding{173}:} At comparable drift magnitude, states with more dispersed critical drift incur greater path deviation when refresh is deferred. Critical Feature Drift Entropy therefore conditions the refresh threshold on tolerance to deferral.}

% ========================================================================
% subsection 3: Drift Continuity Analysis for Adaptive Compensation
% ========================================================================
\subsection{\textbf{Drift Continuity Analysis for Adaptive Compensation}}
\label{sec:analysis_3}

Once a refresh has been requested, a latency spike can leave the robot using $z_k$ for several more steps. During this wait, the edge can reuse $z_k$ or extend the critical component of the last received update, $\Delta z_k^c$ (Sec.~\ref{tex:method_3}). We ask when extending this update reduces mismatch relative to direct reuse. For each observation, we compare the two choices using the fresh latent as the offline reference for critical mismatch and its predicted action as the reference for action error.

To assess whether the last update still provides a useful correction, we compare the current critical feature drift $\mathbf{d}_t$ with the drift $\mathbf{d}_k$ saved when $z_k$ was requested. This reference spans the source observations of $z_{k-1}$ and $z_k$ under $z_{k-1}$. Both vectors use the same feature response coordinates. For nonzero drift vectors, we define \emph{Critical Drift Continuity} as
\begin{equation}
    C_t = \frac{\mathbf{d}_t^\top\mathbf{d}_k}
    {\|\mathbf{d}_t\|_2\,\|\mathbf{d}_k\|_2}.
    \label{eq:analysis_continuity}
\end{equation}
A high $C_t$ means the current feature displacement follows the direction associated with the last received latent update.

Fig.~\ref{fig:analysis3}(a) shows how the value of that update changes within one refresh wait. At the marked step before the turn, $C_t$ is high and extending the update yields 19\% lower critical mismatch than direct reuse. At the marked step after the turn, $C_t$ has dropped and continuing the same correction yields 6\% higher mismatch than reuse. Fig.~\ref{fig:analysis3}(b) tests whether continuity identifies useful steps across refresh waits. At 50\% compensation coverage, selecting steps by $C_t$ reduces action error by 2.7\% relative to reuse, compared with 1.1\% for random selection at the same coverage. This motivates using $C_t$ to select the steps where the edge extrapolates. The ratio $D_t/D_k$ then scales the previous update in the runtime rule.

\vspace{0.5mm}
\noindent\textit{\textbf{Insight \ding{174}:} Latent mismatch can stay small during latency spikes. Extending the last update can reduce mismatch while critical drift stays aligned and increase it after a direction change. Drift continuity gates compensation, and drift magnitude scales the correction.}

\section{\textbf{\textit{AceVLN} Framework}}
\label{tex:method}

\textit{AceVLN} uses critical feature drift to coordinate two runtime decisions in asynchronous cloud--edge VLN (Fig.~\ref{fig:framework}). In \emph{Idle}, the edge uses the drift magnitude and its energy distribution to decide whether to request cloud guidance. In \emph{Pending}, the same drift signal determines whether to correct the cached latent while the request is in flight. The fast system generates actions from the selected latent with its original weights.

\subsection{\textbf{Subspace-Decoupled Latent Mismatch Modeling}}
\label{tex:method_1}

Following Insight~\ding{172}, offline calibration uses the action--latent sensitivities in Sec.~\ref{sec:analysis_1} to obtain an orthonormal latent basis $V_c$ for each backbone. This calibrated basis supplies the runtime critical directions, while $V_c^{\rm loc}$ in Eq.~\eqref{eq:analysis_latent_components} denotes the basis at an individual analysis boundary. With $z\in\mathbb R^{n_z}$ and $h\in\mathbb R^{n_h}$, the first-order feature response to a perturbation $V_c\mathbf u$ at fixed observation is $(\partial h/\partial z)V_c\mathbf u$. The responses to these perturbations identify a subspace in the fused-feature space. In an orthonormal basis $Q_c\in\mathbb R^{n_h\times r}$ of this calibrated response subspace, $P_c=Q_c^\top$ gives the $r$ signed response coordinates used in Eq.~\eqref{eq:analysis_drift}. The edge loads these calibrated bases before navigation and holds them fixed across cache updates. Jacobian and feature-response calculations run offline.

Each request is associated with the source view $o_{s_k}$ from which the cloud generates $z_k$. The edge retains that view until the reply arrives so that drift includes motion between dispatch and installation. When the reply becomes active, the edge caches its source fused feature $h_s=h(o_{s_k},z_k)$. At step $t$, it evaluates $h_t=h(o_t,z_k)$ for the live view and projects $h_t-h_s$ to obtain the directional drift $\mathbf d_t$ and its magnitude $D_t$. Both features are evaluated under $z_k$ before the edge selects the latent for action decoding. The same calibrated response coordinates are used for live and reference drift, so the edge can compare directions across successive updates.

\subsection{\textbf{Drift-Distribution-Guided Latent Refresh Scheduling}}
\label{tex:method_2}

Following Insight~\ding{173}, at each \emph{Idle} step with $D_t>0$, the edge computes the entropy $H_t$ of the drift energy shares using Eq.~\eqref{eq:critical_entropy}. The refresh threshold $\theta(H_t)$ is $\theta_{\rm hi}$ for $H_t<\tau_H$ and $\theta_{\rm lo}$ otherwise, with $0<\theta_{\rm lo}<\theta_{\rm hi}$. This rule lets concentrated drift accumulate further when deferral is better tolerated. At the same $D_t$, dispersed drift can trigger an earlier refresh by selecting the lower threshold. The refresh settings are chosen on development episodes for the desired call rate (Sec.~\ref{sec:experiments_1}).

When $D_t\geq\theta(H_t)$, the edge sends $o_t$, saves $\mathbf d_t$ with the request, and enters \emph{Pending}. At zero drift or below threshold, the edge keeps using $z_k$. The edge allows at most one request in flight and continues to generate actions as it waits. The waiting state enables local compensation without issuing a second cloud request.

\subsection{\textbf{Continuity-Guided Latent Compensation}}
\label{tex:method_3}

Starting with the second cloud reply, the edge pairs each critical latent update $\Delta z_k^c=V_c V_c^\top(z_k-z_{k-1})$ with the reference drift $\mathbf d_k$ saved when $z_k$ was requested. That reference measures the change from $o_{s_{k-1}}$ to $o_{s_k}$ under $z_{k-1}$. Its magnitude $D_k=\|\mathbf d_k\|_2$ sets the scale for extending the associated update.

Following Insight~\ding{174}, while a request is \emph{Pending}, the edge checks drift continuity $C_t$ using Eq.~\eqref{eq:analysis_continuity} when the reference exists and $D_t,D_k>0$. If $C_t\geq\tau_c$, the fast policy uses
\begin{equation}
    \tilde z_t=z_k+
    \frac{\min(D_t,\,\alpha_{\max}D_k)}{D_k}\,\Delta z_k^c .
    \label{eq:latent_compensation}
\end{equation}
Otherwise it uses $z_k$. The gain assumes that drift magnitude and the required correction scale together over a short, aligned segment. Equal drift magnitudes yield unit gain before clipping, so the ratio scales the last update by the relative feature displacement. Clipping before division bounds the correction by $\alpha_{\max}\|\Delta z_k^c\|_2$ for small positive $D_k$. Each step starts from $z_k$, so corrections do not accumulate.

In the execution trace in Fig.~\ref{fig:framework}, drift remains aligned after dispatch and the edge compensates while awaiting the cloud reply. At the turn, drift loses alignment and the policy returns to the cached latent. On receipt of $z_{k+1}$, the edge replaces the cache, recomputes the source feature for the new latent, and stores the latest update and its reference drift. The edge then returns to the \emph{Idle} state.

\section{\textbf{Experiments}}
\label{sec:experiments}

\begin{table}[!t]
\centering
\setlength{\abovecaptionskip}{0.5pt}
\caption{Evaluation of \textit{AceVLN} and baselines~\cite{yang2026tud, liu2026latentbridge, cloudedge} on \textcolor{blue}{\textbf{R2R-CE}}.}
\vspace{-1mm}
\label{tab:projected-r2r}
\fontsize{7.4}{8.2}\selectfont
\setlength{\tabcolsep}{1.35pt}
\renewcommand{\arraystretch}{0.98}
\setlength{\aboverulesep}{0.2ex}
\setlength{\belowrulesep}{0.2ex}
\setlength{\cmidrulesep}{0.15ex}
\setlength{\doublerulesep}{1.4pt}
\newcommand{\AceMetric}[2]{#1\kern.4pt\ensuremath{\scriptstyle#2}}
\newcommand{\AcePair}[2]{#1\kern.75pt{\fontsize{5.7}{7}\selectfont\color{black!52}/}\kern.75pt#2}
% Paint a full-width background behind one data row.
\newcommand{\AceHighlight}[1]{%
  \noalign{\begingroup
    \color{#1}
    \dimen0=\dimexpr\ht\csname @arstrutbox\endcsname+\dp\csname @arstrutbox\endcsname\relax
    \hrule height\dimen0
    \kern-\dimen0
  \endgroup}}
\begin{tabular*}{\columnwidth}{@{\extracolsep{\fill}}l*{3}{c}@{\hspace{5pt}}ccc@{}}
\toprule[.5pt]
\toprule[.28pt]
\multirow{2}{*}{\textbf{Method}} & \multicolumn{3}{c}{\textbf{Navigation}} & \multicolumn{3}{c@{}}{\textbf{Efficiency}} \\
\cmidrule[.26pt](lr){2-4}\cmidrule[.26pt](l){5-7}
 & \AceMetric{NE}{\downarrow} & \AceMetric{SR}{\uparrow} & \AceMetric{SPL}{\uparrow} & \AceMetric{Am}{\downarrow} & \AceMetric{Rsp}{\downarrow} & \AceMetric{FLOPs}{\downarrow} \\
\midrule[.43pt]
\AceHighlight{black!5}
\multicolumn{2}{@{}l}{\textbf{DualVLN}\,\cite{dualvln}} & \multicolumn{5}{r@{}}{\fontsize{7}{8}\selectfont Orin\,/\,Thor (A100 cloud)} \\[.5pt]
Naive & \AcePair{4.48}{4.44} & \AcePair{58.9}{59.4} & \AcePair{51.2}{52.3} & \AcePair{315}{237} & \AcePair{\textbf{193}}{\textbf{113}} & 9450 \\
TUD & \AcePair{4.32}{4.29} & \AcePair{61.0}{61.4} & \AcePair{54.2}{54.8} & \AcePair{278}{199} & \AcePair{194}{\textbf{113}} & 6980 \\
Bridge & \AcePair{4.25}{4.23} & \AcePair{61.8}{62.1} & \AcePair{55.4}{56.0} & \AcePair{263}{182} & \AcePair{198}{115} & 5740 \\
CloudEdge & \AcePair{4.23}{4.20} & \AcePair{62.0}{62.3} & \AcePair{55.7}{56.3} & \AcePair{290}{212} & \AcePair{\textbf{193}}{\textbf{113}} & 7800 \\
\midrule[.35pt]
\AceHighlight{gray!20}
\textbf{\textit{AceVLN}} & \AcePair{\textbf{4.09}}{\textbf{4.07}} & \AcePair{\textbf{63.8}}{\textbf{64.0}} & \AcePair{\textbf{57.8}}{\textbf{58.0}} & \AcePair{\textbf{256}}{\textbf{172}} & \AcePair{210}{124} & \textbf{4620} \\
\midrule[.28pt]
\multicolumn{2}{@{}l}{\textbf{NavDP}\,\cite{internrobotics2025navdp}} \\[.5pt]
Naive & \AcePair{4.72}{4.67} & \AcePair{58.1}{58.7} & \AcePair{50.3}{51.4} & \AcePair{428}{281} & \AcePair{\textbf{293}}{\textbf{144}} & 9950 \\
TUD & \AcePair{4.54}{4.50} & \AcePair{60.3}{60.8} & \AcePair{53.3}{54.0} & \AcePair{387}{239} & \AcePair{294}{\textbf{144}} & 7480 \\
Bridge & \AcePair{4.46}{4.42} & \AcePair{61.3}{61.7} & \AcePair{54.8}{55.4} & \AcePair{370}{220} & \AcePair{298}{146} & 6240 \\
CloudEdge & \AcePair{4.41}{4.38} & \AcePair{61.7}{62.1} & \AcePair{55.2}{55.8} & \AcePair{400}{253} & \AcePair{\textbf{293}}{\textbf{144}} & 8300 \\
\midrule[.35pt]
\AceHighlight{gray!20}
\textbf{\textit{AceVLN}} & \AcePair{\textbf{4.25}}{\textbf{4.23}} & \AcePair{\textbf{63.7}}{\textbf{64.0}} & \AcePair{\textbf{57.5}}{\textbf{57.8}} & \AcePair{\textbf{359}}{\textbf{204}} & \AcePair{308}{151} & \textbf{5090} \\
\bottomrule[.28pt]
\bottomrule[.5pt]
\end{tabular*}
\vspace{-4mm}
\end{table}

% ===========================================================================
% subsection 1: Setup
% ===========================================================================
\subsection{\textbf{Experimental Setup}}
\label{sec:experiments_1}

\noindent\textbf{Benchmarks and Deployment.}
We evaluate \textit{AceVLN} with DualVLN~\cite{dualvln} and InternVLA-N1 with NavDP~\cite{internrobotics2025navdp} on R2R-CE~\cite{krantz2020beyond} and English RxR-CE~\cite{ku2020room} (val-unseen). Simulated navigation uses a remote A100 (S2) and Jetson AGX Orin or Thor (S1). Methods share episodes, controls, and the native link (nominal RTT 172.3\,ms).

\noindent\textbf{Baselines and Metrics.}
\emph{Naive} refreshes periodically, \emph{TUD}~\cite{yang2026tud} uses action-plan disagreement, \emph{Bridge}~\cite{liu2026latentbridge} predicts latent deltas, and \emph{CloudEdge}~\cite{cloudedge} trains for delay tolerance. We fix refresh settings on shared development episodes for each method's target call rate.
We report NE, SR, SPL, and RxR-CE nDTW. Over $N$ new decisions, $\mathrm{Am}=(T_{\rm edge}+T_{\rm cloud})/N$ includes monitoring, prediction, and compensation. Edge is $T_{\rm edge}/N$. Rsp times observation-to-command after initialization, including blocking waits excluded from Am (mean in tables, P95 under stress). FLOPs sum edge and cloud GFLOPs/decision. S2 counts calls/100 decisions.

\begin{table}[!b]
\centering
\vspace{-4mm}
\setlength{\abovecaptionskip}{0.5pt}
\caption{Evaluation of \textit{AceVLN} and baselines~\cite{yang2026tud,liu2026latentbridge,cloudedge} on \textcolor{blue}{\textbf{RxR-CE}}.}
\vspace{-1mm}
\label{tab:projected-rxr}

\fontsize{7.4}{8.2}\selectfont
\setlength{\tabcolsep}{1.35pt}
\renewcommand{\arraystretch}{0.98}
\setlength{\aboverulesep}{0.2ex}
\setlength{\belowrulesep}{0.2ex}
\setlength{\cmidrulesep}{0.15ex}
\setlength{\doublerulesep}{1.4pt}
\newcommand{\AceMetric}[2]{#1\kern.4pt\ensuremath{\scriptstyle#2}}
\newcommand{\AcePair}[2]{#1\kern.75pt{\fontsize{5.7}{7}\selectfont\color{black!52}/}\kern.75pt#2}
% Paint a full-width background behind one data row.
\newcommand{\AceHighlight}[1]{%
  \noalign{\begingroup
    \color{#1}
    \dimen0=\dimexpr\ht\csname @arstrutbox\endcsname+\dp\csname @arstrutbox\endcsname\relax
    \hrule height\dimen0
    \kern-\dimen0
  \endgroup}}
\begin{tabular*}{\columnwidth}{@{\extracolsep{\fill}}l*{4}{c}@{\hspace{5pt}}ccc@{}}
\toprule[.5pt]
\toprule[.28pt]
\multirow{2}{*}{\textbf{Method}} & \multicolumn{4}{c}{\textbf{Navigation}} & \multicolumn{3}{c@{}}{\textbf{Efficiency}} \\
\cmidrule[.26pt](lr){2-5}\cmidrule[.26pt](l){6-8}
 & \AceMetric{NE}{\downarrow} & \AceMetric{SR}{\uparrow} & \AceMetric{SPL}{\uparrow} & \AceMetric{nDTW}{\uparrow} & \AceMetric{Am}{\downarrow} & \AceMetric{Rsp}{\downarrow} & \AceMetric{FLOPs}{\downarrow} \\
\midrule[.43pt]
\AceHighlight{black!5}
\multicolumn{2}{@{}l}{\textbf{DualVLN}\,\cite{dualvln}} & \multicolumn{6}{r@{}}{\fontsize{7}{8}\selectfont Orin\,/\,Thor (A100 cloud)} \\[.5pt]
Naive & \AcePair{5.26}{5.20} & \AcePair{53.6}{54.3} & \AcePair{42.0}{43.1} & \AcePair{63.2}{64.0} & \AcePair{354}{276} & \AcePair{\textbf{193}}{\textbf{113}} & 11970 \\
TUD & \AcePair{5.00}{4.96} & \AcePair{56.5}{57.0} & \AcePair{46.0}{46.7} & \AcePair{65.7}{66.3} & \AcePair{307}{228} & \AcePair{194}{\textbf{113}} & 8830 \\
Bridge & \AcePair{4.84}{4.81} & \AcePair{58.4}{58.8} & \AcePair{48.3}{48.9} & \AcePair{67.9}{68.3} & \AcePair{290}{209} & \AcePair{198}{115} & 7480 \\
CloudEdge & \AcePair{4.89}{4.84} & \AcePair{57.8}{58.4} & \AcePair{47.7}{48.5} & \AcePair{67.3}{67.9} & \AcePair{327}{248} & \AcePair{\textbf{193}}{\textbf{113}} & 10170 \\
\midrule[.35pt]
\AceHighlight{gray!20}
\textbf{\textit{AceVLN}} & \AcePair{\textbf{4.66}}{\textbf{4.62}} & \AcePair{\textbf{60.5}}{\textbf{60.9}} & \AcePair{\textbf{50.7}}{\textbf{51.1}} & \AcePair{\textbf{69.3}}{\textbf{69.6}} & \AcePair{\textbf{275}}{\textbf{190}} & \AcePair{211}{124} & \textbf{5810} \\
\midrule[.28pt]
\multicolumn{2}{@{}l}{\textbf{NavDP}\,\cite{internrobotics2025navdp}} \\[.5pt]
Naive & \AcePair{5.42}{5.36} & \AcePair{51.4}{52.1} & \AcePair{40.5}{41.8} & \AcePair{62.1}{62.9} & \AcePair{471}{323} & \AcePair{\textbf{293}}{\textbf{144}} & 12470 \\
TUD & \AcePair{5.22}{5.16} & \AcePair{54.2}{54.8} & \AcePair{43.9}{44.7} & \AcePair{64.8}{65.5} & \AcePair{418}{270} & \AcePair{294}{\textbf{144}} & 9330 \\
Bridge & \AcePair{5.04}{4.98} & \AcePair{56.1}{56.7} & \AcePair{46.4}{47.2} & \AcePair{66.9}{67.5} & \AcePair{400}{249} & \AcePair{298}{146} & 7980 \\
CloudEdge & \AcePair{4.99}{4.94} & \AcePair{56.7}{57.1} & \AcePair{47.0}{47.9} & \AcePair{67.4}{68.0} & \AcePair{440}{293} & \AcePair{\textbf{293}}{\textbf{144}} & 10670 \\
\midrule[.35pt]
\AceHighlight{gray!20}
\textbf{\textit{AceVLN}} & \AcePair{\textbf{4.76}}{\textbf{4.72}} & \AcePair{\textbf{59.0}}{\textbf{59.5}} & \AcePair{\textbf{49.7}}{\textbf{50.2}} & \AcePair{\textbf{68.9}}{\textbf{69.4}} & \AcePair{\textbf{380}}{\textbf{224}} & \AcePair{309}{151} & \textbf{6270} \\
\bottomrule[.28pt]
\bottomrule[.5pt]
\end{tabular*}
\end{table}

% ===========================================================================
% subsection 2: End-to-End Results
% ===========================================================================
\subsection{\textbf{End-to-End Results}}

Tables~\ref{tab:projected-r2r} and~\ref{tab:projected-rxr} report R2R-CE and RxR-CE results in Orin/Thor order.
Across both backbones and edge platforms, \textit{AceVLN} improves SR by 1.7--2.4 percentage points and SPL by 1.7--2.7 points over the best baseline per setting. It cuts total FLOPs by 18.4--22.3\% and Am by 2.7--10.0\% relative to \emph{Bridge}, the lowest-Am baseline.

Against \emph{Bridge}, mean Rsp increases by 5--13\,ms. Fewer cloud calls offset the added edge work, cutting Am by 7\,ms on DualVLN (R2R-CE, Orin) and 25\,ms on NavDP (RxR-CE, Thor).

\begin{table}[t]
\centering
\setlength{\abovecaptionskip}{0.5pt}
\caption{RxR-CE ablation on DualVLN~\cite{dualvln} with Orin.}
\vspace{-1mm}
\label{tab:acevln-ablation}
\fontsize{8}{8.9}\selectfont
\setlength{\tabcolsep}{.85pt}
\setlength{\aboverulesep}{.45pt}
\setlength{\belowrulesep}{.45pt}
\renewcommand{\arraystretch}{1}
\newcommand{\ablmetric}[2]{\textbf{#1}\kern.45pt\ensuremath{\scriptstyle#2}}
\begin{tabular*}{\columnwidth}{@{\extracolsep{\fill}}l*{5}{r}@{\hspace{5pt}}*{4}{r}@{}}
\specialrule{.5pt}{0pt}{0pt}
\specialrule{.28pt}{1.4pt}{.8pt}
 & \multicolumn{5}{c}{\textbf{Fixed S2 rate (10.0)}} & \multicolumn{4}{c@{}}{\textbf{Periodic SR target}} \\
\cmidrule[.26pt](lr){2-6}\cmidrule[.26pt](l){7-10}
\multirow{2}{*}{\textbf{Variant}} & Instant & Deployed
 & \multirow{2}{*}{\ablmetric{Edge}{\downarrow}} & \multirow{2}{*}{\ablmetric{Am}{\downarrow}}
 & \multirow{2}{*}{\ablmetric{Rsp}{\downarrow}} & \multirow{2}{*}{\textbf{SR}}
 & \multirow{2}{*}{\ablmetric{S2}{\downarrow}} & \multirow{2}{*}{\ablmetric{Am}{\downarrow}}
 & \multirow{2}{*}{\ablmetric{Rsp}{\downarrow}} \\
 & \textbf{SR}\,{\scriptsize\itshape ref.} & \ablmetric{SR}{\uparrow} & & & & & & & \\
\specialrule{.43pt}{.65pt}{.7pt}
Periodic & 56.0 & 52.8 & \textbf{188} & \textbf{257} & \textbf{193} & 52.8 & 10.0 & 257 & \textbf{193} \\
Periodic + gate & 56.0 & 55.5 & 206 & 275 & 211 & 53.3 & 9.1 & 268 & 210 \\
\specialrule{.25pt}{.45pt}{.45pt}
Full drift & 58.3 & 55.1 & 205 & 274 & 210 & 53.1 & 9.0 & 267 & 210 \\
Critical drift & 60.0 & 56.8 & 205 & 274 & 210 & 53.2 & 8.3 & 262 & 209 \\
+ Entropy & \textbf{61.0} & 57.8 & 205 & 274 & 210 & 53.0 & 7.8 & 259 & 210 \\
+ Ungated comp. & \textbf{61.0} & 59.1 & 205 & 275 & 210 & 53.1 & 7.5 & 257 & 210 \\
\textbf{\textit{AceVLN}} & \textbf{61.0} & \textbf{60.5} & 206 & 275 & 211 & 53.2 & \textbf{7.0} & \textbf{254} & 210 \\
\specialrule{.28pt}{.9pt}{0pt}
\specialrule{.5pt}{1.4pt}{0pt}
\end{tabular*}
\vspace{-3mm}
\end{table}

% ===========================================================================
% subsection 3: Ablation Study
% ===========================================================================
\subsection{\textbf{Ablation Study}}

In Table~\ref{tab:acevln-ablation}, Instant SR uses each requested latent at dispatch. Deployed SR includes the refresh wait.

At 10.0 S2 calls per 100 decisions, periodic refresh reaches 52.8\% SR and 55.5\% with gated compensation. Full-space drift reaches 55.1\%, and restricting it to the Critical Subspace adds 1.7 points (56.8\%). Entropy scheduling reaches 57.8\%. Ungated compensation raises SR to 59.1\%, and continuity gating raises it to 60.5\%.

Using the periodic policy's 52.8\% SR as the minimum target, \textit{AceVLN} cuts S2 calls by 30\% (10.0 to 7.0) and lowers Am from 257 to 254\,ms.

\begin{figure}[!t]
    \centering
    \includegraphics[width=3.3in]{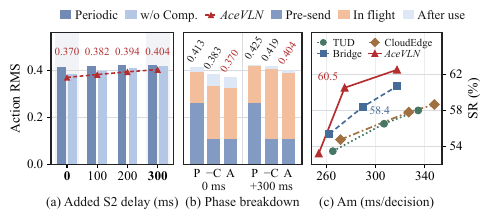}
    \vspace{-3mm}
    \caption{Action deviation and compute--accuracy trade-offs.}
    \vspace{-4mm}
    \label{fig:refresh-delay}
\end{figure}

\begin{table}[t]
\centering
\setlength{\abovecaptionskip}{0.5pt}
\caption{S2 cost trade-offs on R2R-CE (DualVLN, Orin).}
\vspace{-1mm}
\label{tab:s2-acceleration}
\fontsize{8}{9.6}\selectfont
\setlength{\tabcolsep}{1.15pt}
\renewcommand{\arraystretch}{1.02}
\begin{tabular*}{\columnwidth}{@{\extracolsep{\fill}}lrrrrrrr@{}}
\specialrule{.5pt}{0pt}{0pt}
\specialrule{.28pt}{1.4pt}{1pt}
 & \multicolumn{2}{c}{\textbf{Policy}}
 & \multicolumn{3}{c}{\textbf{Am}\,$\downarrow$ (ms/decision)}
 & \multicolumn{2}{c}{\textbf{SR}\,$\uparrow$ (\%)} \\
\cmidrule[.27pt](lr){2-3}\cmidrule[.27pt](lr){4-6}\cmidrule[.27pt](l){7-8}
\textbf{Method} & S2/100 & Edge & 637\,ms & 419\,ms & 350\,ms & 637\,ms & 419\,ms \\
\specialrule{.43pt}{.6pt}{.8pt}
Naive     & 20 & 188.0 & 315.40 & 271.80 & 258.00 & 58.9 & 59.2 \\
TUD       & 14 & 189.0 & 278.18 & 247.66 & 238.00 & 61.0 & 60.8 \\
CloudEdge & 16 & 188.0 & 289.92 & 255.04 & 244.00 & 62.0 & 61.4 \\
Bridge    & 11 & 193.3 & 263.41 & 239.43 & \textbf{231.84} & 61.8 & 61.4 \\
\textbf{\textit{AceVLN}} & 8 & 205.2 & \textbf{256.20} & \textbf{238.76}
 & 233.24 & \textbf{63.8} & \textbf{63.0} \\
\specialrule{.32pt}{.8pt}{.8pt}
\multicolumn{3}{@{}l}{\textit{AceVLN} $-$ Bridge}
 & $-7.21$ & $-0.67$ & $+1.40$ & $+2.0$ & $+1.6$ \\
\specialrule{.28pt}{.8pt}{0pt}
\specialrule{.5pt}{1.4pt}{0pt}
\end{tabular*}
\vspace{-4mm}
\end{table}

% ===========================================================================
% subsection 4: Runtime and Deployment Analysis
% ===========================================================================
\subsection{\textbf{Runtime and Deployment Analysis}}

\noindent\textbf{In-Flight Correction and Compute--Accuracy Trade-offs.}
Fig.~\ref{fig:refresh-delay}(a,b) adds 0--300\,ms to S2 result delivery at fixed compute time. \textit{AceVLN}'s action RMS rises from 0.370 to 0.404. At +300\,ms, RMS is 0.425 for \emph{Periodic} and 0.419 with compensation disabled.
Sweeping three S2 call rates on RxR-CE in Fig.~\ref{fig:refresh-delay}(c), \textit{AceVLN} reaches 60.5\% SR at 275\,ms Am versus \emph{Bridge}'s 58.4\% at 290\,ms, keeping a 1.8-point lead near 317\,ms.

Table~\ref{tab:s2-acceleration} varies S2 compute per call at fixed call rates and edge times. At 637 and 419\,ms, \textit{AceVLN} saves 7.21 and 0.67\,ms in Am and gains 2.0 and 1.6 SR points over \emph{Bridge}, respectively. At 350\,ms, \emph{Bridge} saves 1.40\,ms in Am.

\begin{figure}[!b]
    \centering
    \vspace{-5mm}
    \includegraphics[width=3.3in]{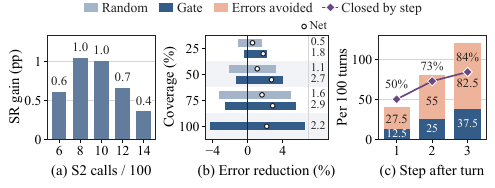}
    \vspace{-3mm}
    \caption{Refresh budgets and compensation behavior.}
    \label{fig:refresh-gate}
\end{figure}

\begin{figure}[!b]
    \centering
    \vspace{-4mm}
    \includegraphics[width=3.3in]{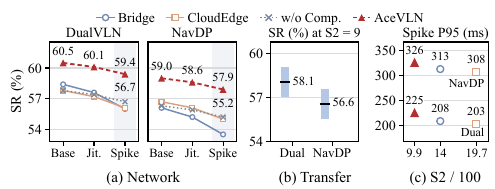}
    \vspace{-3mm}
    \caption{Network stress and calibration transfer.}
    \label{fig:deployment-studies}
\end{figure}

\noindent\textbf{Refresh Scheduling and Continuity Gating.}
In Fig.~\ref{fig:refresh-gate}(a), entropy scheduling peaks at a 1.0-point SR gain under 8--10 S2 calls per 100 decisions.
In Fig.~\ref{fig:refresh-gate}(b), continuity gating raises net error reduction at 50\% coverage from 1.1\% (\emph{Random}) to 2.7\% and reaches 2.9\% at 75\%, surpassing always-on 100\% compensation (2.2\%).
After turns (Fig.~\ref{fig:refresh-gate}(c)), 84\% of open gates close within three steps, avoiding 82.5 harmful updates per 100 turns.

\noindent\textbf{Network Stress and Calibration Transfer.}
Jitter adds 5--195\,ms per 100\,ms slot and spikes add 1\,s for 2\,s every 20\,s. Both add 100\,ms to RTT on average (Fig.~\ref{fig:deployment-studies}(a), one request in flight at most). Under spikes, enabling compensation raises SR by 2.7 points on each backbone. The transfer panel shows RxR-CE reference SRs of 58.1\% (DualVLN) and 56.6\% (NavDP) at nine calls per 100 decisions, each with a $\pm1$-point band (Fig.~\ref{fig:deployment-studies}(b)). Against \emph{Bridge} under spikes, \textit{AceVLN} cuts S2 calls from 14.0 to 9.9 per 100 decisions and adds 17\,ms (DualVLN) or 13\,ms (NavDP) to P95 Rsp (Fig.~\ref{fig:deployment-studies}(c)).

\section{\textbf{Conclusion}}
\label{sec:conclusion}

We present \textit{AceVLN}, a training-free framework for efficient and reliable cloud--edge deployment of streaming VLN models. 
By exploiting critical subspace mismatch, adaptive latent refresh, and continuity-guided compensation, \textit{AceVLN} reduces cloud computation while maintaining navigation performance and improving robustness to latency spikes. 
Extensive experiments demonstrate the effectiveness of runtime latent management for asynchronous embodied navigation.

% ==================================================================
% reference
% ==================================================================
\clearpage
\twocolumn
\bibliographystyle{IEEEtran}
\IEEEtriggeratref{15}
\bibliography{ref/references}

\end{document}